\documentclass[traditabstract]{aa}

\usepackage{amsmath}
\usepackage{amssymb}
\usepackage{graphicx}
\usepackage{txfonts}
\usepackage{dcolumn}
\usepackage{natbib}
\usepackage{url}
\usepackage{color}
\usepackage{units}
\bibpunct{(}{)}{;}{a}{}{,}

\newcommand{\diff}{{\textrm{d}}}

\newcommand{\mrm}[1]{{\mathrm{#1}}}

\begin{document}

\title{Convective settling in main sequence stars: Li and Be depletion}

\author{R.\ Andr\'assy, H.\ C.\ Spruit}

\institute{Max Planck Institute for Astrophysics, Karl-Schwarzschildstr. 1, 85748 Garching, Germany}

\date{Received ; accepted }

\abstract{The process of convective settling is based on the assumption that a small fraction of the low-entropy downflows sink from the photosphere down to the bottom of the star's envelope convection zone retaining a substantial entropy contrast. We have previously shown that this process could explain the slow Li depletion observed in the Sun. We construct a parametric model of convective settling to investigate the dependence of Li and Be depletion on stellar mass and age. Our model is generally in good agreement with the Li abundances measured in open clusters and solar twins, although it seems to underestimate the Li depletion in the first $\sim 1$\,Gyr. The model is also compatible with the Be abundances measured in a sample of field stars.}

\maketitle
%

\section{Introduction}

Low-mass, main sequence stars, with their deep convective envelopes, are astrophysical laboratories that allow us to investigate weak mixing processes under stellar conditions. The small distance between the convection zone proper and the Li-burning layer in these stars makes the surface abundance of Li a sensitive indicator of any mixing process that may be at work just below the convection zone. The surface abundance of Be, which is burnt at a somewhat higher temperature, provides an additional constraint on the extent of this mixing.

The abundance of Li in the Sun is more than two orders of magnitude lower than  in  meteorites \citep{GreensteinRichardson51, AndersGrevesse89, AsplundEtal09}. Lithium abundances observed in open clusters \citep{Herbig65, Danziger69, Zappala72} and solar analogues and twins \citep{BaumannEtal10, MelendezEtal10, MonroeEtal13, MelendezEtal14} of different ages show that Li depletion takes place systematically on the main sequence, although sufficiently cool stars also deplete significant amounts of Li  in the pre-main-sequence phase \citep{MagazzuRebolo89, MartinEtal94, SestitoRandich05}. The situation is further complicated by the observed spread in the Li abundances in cool stars of the same age, composition, and effective temperature \citep{DuncanJones83, SoderblomEtal93a, SoderblomEtal93b, Garcia-LopezEtal94, JonesEtal99}. At the same effective temperature, fast rotators are more Li-rich than the slow ones in the Pleiades and in $\alpha$~Per \citep{SoderblomEtal93a, BalachandranEtal11}. Lithium distributions in open clusters also contain a prominent feature in the temperature range between 6200\,K and 7000\,K, the so-called Li dip, in which the Li abundance drops to very low levels \citep{BoesgaardTripicco86}.

The abundance of Be is notoriously hard to measure. Its value in the Sun is equal to the meteoritic value within 0.1\,--\,0.2\,dex \citep{BalachandranBell98, Asplund04}. Beryllium studies in open clusters have mostly been focused on stars hotter than 6000\,K with the general conclusion that the Li dip stars are also Be-deficient \citep{Boesgaard76, StephensEtal97} and that there is a Be-Li correlation on the cool side of the dip \citep{BoesgaardEtal01, BoesgaardEtal04a, BoesgaardEtal04b}. Cool ($T_\text{eff} \lesssim 6000\,\text{K}$), main-sequence field stars show significant Be depletion \citep{SantosEtal04, Delgado-MenaEtal12}, although much smaller than that of Li.

The standard solar model predicts negligible Li and Be depletion on the main sequence \citep{SchwarzschildEtal57, WeymannSears65, Pinsonneault97, SchlattlWeiss99}. To explain the observations quoted above, we have to include an additional, slow mixing process into the standard models of low-mass stars. It should mix the material from the convection zone proper down to the Li- and Be-burning layers and the mixing rate should quickly decrease with depth.

Several candidate processes have been proposed (see \citeauthor{Chaboyer98}~\citeyear{Chaboyer98} for a more comprehensive review). The most popular is rotation-induced mixing \citep{Zahn92, CharbonnelEtal94, MeynetMaeder97, Maeder97}, especially for its ability to naturally explain the observed Li spread in stars of the same mass and age by ascribing it to the wide range of rotation periods stars are born with. One would expect the depletion rate due to rotation-induced mixing to increase with decreasing rotation period (the opposite to what is observed, see above), but it is possible to reverse this trend by assuming that the mixing-length parameter $\alpha_{MLT}$ decreases with decreasing rotation period \citep{SomersPinsonneault15}.

Alternatively, the mixing might be caused by some form of convective overshooting, usually considered in the diffusion approximation suggested by \citet{Freytag96}. If this is the case, the overshooting parameter $f_{ov}$ must take a different value on the pre-main sequence than on the main sequence \citep{SchlattlWeiss99}. Mixing induced by internal gravity waves has also been employed to explain the Li- and Be-depletion patterns in low-mass stars, including the Li dip \citep{Press81, Garcia-LopezSpruit91, Montalban94, Schatzman96}. In this paper we focus on the process we call ``convective settling'' proposed by \citet{Spruit97} and elaborated in \citet[][hereafter Paper~I]{AndrassySpruit13}.

Envelope convection zones are dominated by large-scale downflows spanning the whole convection zone \citep[see][]{NordlundStein97, TrampedachEtal14}. They are generated by the strong radiative cooling in the photosphere, which makes their initial entropy much lower than that of the nearly isentropic upflow. As they sink they merge, entrain mass from the hot upflow, and are heated by radiative diffusion on small scales. If a small fraction ($\approx 10^{-7}$, see Paper~I) of the photospheric downflows retain a substantial entropy deficit until they have arrived at the base of the convection zone, they will continue sinking until each of them has settled on its level of neutral buoyancy, hence the name ``convective settling''. The entropy the downflows start with is low enough for them to reach the Li-burning layer. In general, there will be a broad distribution of entropy contrasts at the base of the convection zone, spanning from the low values predicted by the mixing-length theory (MLT) up to the highest values the downflows start with in the photosphere. They will settle at a range of depths, and mass conservation will enforce an upflow carrying the Li- and Be-depleted material back to the convection zone, reducing the surface abundances.

The envelope convection problem is too difficult even for the state-of-the-art numerical simulations. Hence, the distribution mentioned above can only be parametrised. In Paper~I, we show that a power-law parametrisation leads to a model that can explain the slow, main-sequence Li depletion in the Sun without changing the thermal stratification enough  to come into conflict with the results of helioseismology. The simplified model presented in Paper~I, however, is based on an approximate, non-evolving model of the Sun and it does not utilise the Be constraint. In this paper, we construct a model of the convective settling process that takes stellar evolution into account and applies both the Li and Be constraints. We also extend the computation to a range of stellar masses (from $0.8\,M_\odot$ to $1.2\,M_\odot$) in order to compare predictions of the model with abundance measurements in open clusters, solar twins, and field stars. We do not attempt to reproduce the Li/Be dip, which is probably due to a separate process \citep{Garcia-LopezSpruit91}.

\section{Model}
\subsection{Overview}
\label{sect:overview}

We construct a two-component, kinematic model of the convective settling process. The first component is an ensemble of downflows leaving the convection zone and sinking towards the deeper layers. They are created by the rapid cooling in the photosphere, and the entrainment and heating processes  change their distribution on their way through the convection zone. We do not model these processes. Instead, we parametrise their output by a distribution of a mass flow rate\footnote{We use the term mass flux (measured in $\text{g\,cm}^{-2}\,\text{s}^{-1}$ in cgs units) in Paper I. In this paper, we work in spherical geometry and the mass flow rate (measured in $\text{g}\,\text{s}^{-1}$ in cgs units) becomes a more convenient quantity.} in the downflows with respect to their entropy contrast and model how they settle below the convection zone. Each downflow settles at the point where its entropy equals that of its surroundings, i.e.\ when it becomes neutrally buoyant. The other component of the model is an upflow due to mass conservation, the strength of which at any given depth depends on the total settling rate below that depth. The upflow advects the Li- and Be-depleted gas back to the convection zone.

The distribution of the mass flow rate is a central element in our model. The downflows at the base of an envelope convection zone are characterised by a wide range of entropy contrasts $\delta s$ with respect to the isentropic upflow. This range can be split into two parts. One part represents the well-mixed material and the fluctuations created \emph{\emph{locally}} by turbulent convection. Their typical amplitude $(\delta s)_\text{typ}$ is very small owing to the high efficiency of convection deep in the stellar interior. Therefore, they settle in a very thin layer below the convection zone. Downflows with an entropy contrast significantly greater than $(\delta s)_\text{typ}$ correspond to the incompletely mixed remnants of the photospheric downflows. They span several orders of magnitude in $\delta s$, up to $(\delta s)_\text{max}$, which is the maximum entropy contrast reached by the downflows just below the photosphere. The coolest of these downflows settle deep below the convection zone. We ignore the well-mixed component in our model and parametrise the distribution of the mass flow rate carried by the incompletely mixed downflows. We do not model the entrainment and heating processes explicitly; the mass flow rate as a function of the contrast value as parametrised in the model refers to the point where the downflow ultimately settles. We call this distribution \emph{\emph{the settling rate distribution}}.

We assume that the settling rate distribution is a power law defined over the range of downflow entropy contrasts \mbox{$(\delta s)_\text{min} \le \delta s \le (\delta s)_\text{max}$}. The minimal entropy contrast considered, $(\delta s)_\text{min}$, corresponds to the transition point to the range of the well-mixed downflows, thus we set $(\delta s)_\text{min} = (\delta s)_\text{typ}$. We estimate $(\delta s)_\text{typ}$ using the MLT and extract $(\delta s)_\text{max}$ from radiation-hydrodynamic simulations of stellar photospheres, which are readily available today (see Sect.~\ref{sect:mathematical_formulation}).

In the settling paradigm, the slope $\beta$ of the settling rate distribution should be a characteristic value resulting from the physics of the entrainment and heating processes in the convection zone and in the settling region. Therefore, we keep this value -- albeit unknown -- constant for all stars and all their evolutionary stages,  and investigate what influence $\beta$ has on the results.

The total mass flow rate $\dot{M}$ of the distribution is given by the mass flow rate that is leaving the photosphere, but it is considerably modified by the entrainment and heating processes in the convection zone and in the settling region. One could argue that the entropy of a downflow depends on the radiative heat exchange on small scales between the core of the downflow and the entrained material, which takes place on the downflow's way from the photosphere to its settling point. The relative importance of radiation in the convection zone is parametrised through the value of $(\delta s)_\text{typ}$. The stronger the radiative heating of the downflows, the higher the value of $(\delta s)_\text{typ}$. On the other hand, the stronger the radiative heating, the smaller the fraction of the cold downflows that reach the settling region. To take this into account we scale $\dot{M}$ both in proportion to the mass downflow rate in the photosphere and in inverse proportion to $(\delta s)_\text{typ}$. We use this scaling so as to include the qualitative effect of the heating process, although we realise that more parameters are likely to play a role and the dependence is much more complicated in reality. The constant of proportionality in the scaling is adjusted until the model reproduces the observed Li depletion in the Sun for a given \mbox{value of $\beta$}.

The convective settling process, depending on its strength, can change the thermal stratification below the convection zone. In Paper~I, we  quantified this effect and showed it to be negligible in a solar model calibrated to reproduce the Li depletion observed in the solar photosphere if the input settling rate distribution is not too steep ($\beta \lesssim 2.5$). In this work, we neglect the influence of convective settling on the thermal stratification and compute Li and Be depletion using a few pre-computed stellar-evolution models.

\subsection{Stellar models}
\label{sect:stellar_models}

We consider the main-sequence evolution of five solar-composition stars with masses of $0.8\,M_\odot$, $0.9\,M_\odot$, $1.0\,M_\odot$, $1.1\,M_\odot$, and $1.2\,M_\odot$. The models have been computed with the stellar-evolution code GARSTEC \citep{WeissSchlattl08}, neglecting the processes of convective overshooting and gravitational settling (also called sedimentation or diffusion). Since the code does not output the stratification of the specific entropy $s$ in the star, we compute an approximation to it from the stratification of the temperature $T$ and pressure $p$ using the ideal-gas expression
\begin{equation}
s = \mathbb{R} \ln\left(\frac{T^{5/2}}{p}\right),
\label{eq:entropy}
\end{equation}
where $\mathbb{R} = k_\text{B}/(\mu\,m_\text{u})$ is the gas constant including the mean molecular weight $\mu$, $k_\text{B}$ the Boltzmann constant, and $m_\text{u}$ the atomic mass unit. The mean molecular weight is constant in the region we are interested in. Equation~\ref{eq:entropy} assumes a constant level of ionisation, which is a good assumption at the temperatures and densities prevailing below the convection zones of the stars considered.

\subsection{Mathematical formulation}
\label{sect:mathematical_formulation}

\subsubsection{Parametric model}

The settling rate $\dot{m}$ in the downflows is distributed as
\begin{equation}
\diff\dot{m} = \dot{M} f(\delta s)\,\diff \delta s
,\end{equation}
where $\dot{M} \ge 0$ is the total settling rate, and $\delta s = s_0 - s_\mrm{d}$ is the entropy contrast with $s_0$ the entropy of the stratification at the bottom of the convection zone and $s_\mrm{d}$ the entropy of the downflow. As discussed in Sect.~\ref{sect:overview}, the distribution function $f(\delta s)$ is assumed to be a power law,
\begin{equation}
f(\delta s) = 
\begin{cases}
N\,\left(\frac{\delta s}{(\delta s)_\text{min}}\right)^{-\beta} & \text{for } (\delta s)_\text{min} \le \delta s \le (\delta s)_\text{max}, \\[0.25cm]
0 & \text{otherwise},
\end{cases}
\label{eq:distribution_function}
\end{equation}
where $N$ is a normalisation factor, $(\delta s)_\text{min} > 0$ and \mbox{$(\delta s)_\text{max} > 0$} are the bounds, and $\beta > 0$ is the slope of the distribution. We require  $\int_{-\infty}^\infty f(\delta s)\,\diff\delta s = 1$ , so that
\begin{equation}
N = 
\begin{cases}
\frac{\beta - 1}{(\delta s)_\text{min}} \left[1 - \left(\frac{(\delta s)_\text{max}}{(\delta s)_\text{min}}\right)^{-(\beta - 1)}\right]^{-1} & \text{for } \beta \neq 1, \\[0.25cm]
\left[(\delta s)_\text{min} \ln\left(\frac{(\delta s)_\text{max}}{(\delta s)_\text{min}}\right)\right]^{-1} & \text{for } \beta = 1.
\end{cases}
\label{eq:normalisation_factor}
\end{equation}

A downflow of entropy $s_\text{d} = s_0 - \delta s$ reaches neutral buoyancy and settles down at the point where the entropy of the surrounding stratification $s$ equals $s_\text{d}$. Mass conservation requires the upward mass flow rate at this point to be $\dot{M} F(s)$, where $F(s)$ is the fraction of downflows that settle below\footnote{Since specific entropy decreases with increasing pressure in a stable thermal stratification.} this point and is given by the cumulative distribution function
\begin{equation}
F(s) = \int\limits_{-\infty}^s f(s_0 - s^\prime)\ \diff s^\prime.
\label{eq:cumulative_distribution_function}
\end{equation}

\subsubsection{Parameter scaling}
\label{sect:parameter_scaling}

As described above, there are four parameters to be specified: $(\delta s)_\text{min}$, $(\delta s)_\text{max}$, $\dot{M}$, and $\beta$. In Paper~I, we used fixed values because we only modelled one star, the evolution of which was also neglected. Now, we intend to model a range of different stars and follow their evolution, hence we have to adapt the parameter values to the changing physical conditions in the convection zone. The only exception is $\beta$, which is held constant (see Sect.~\ref{sect:overview}).

The entropy contrast of the coldest downflow in the ensemble, $(\delta s)_\text{max}$, is given by the maximum entropy contrast that the cooling process in the photosphere can create. This value reaches a well-defined maximum (in a time-averaged sense) just below the photosphere and can be extracted from radiation-hydrodynamic simulations of stellar photospheres. We use the grid of models computed by \citet{MagicEtal13} and approximate the dependence of $(\delta s)_\text{max}$ on the effective temperature $T_\text{eff}$ of the star and on its surface gravity $\log g$ by the fitting function
\begin{align}
\log\,(\delta s)_\text{max} = a_0 + a_1 x + a_2 y,
\label{eq:ds_max}
\end{align}
where $x = (T_\text{eff} - 5777)/1000$, $y = \log g - 4.44$, $a_0 = 8.164$, $a_1 = 0.491$, and $a_2 = -0.461$ with cgs units assumed throughout.

We set the lowest entropy contrast considered, $(\delta s)_\text{min}$, equal to a typical entropy contrast $(\delta s)_\mrm{typ}$ predicted by the MLT (see Sect.~\ref{sect:overview}). We use the MLT formulation of \citet{KippenhahnEtal12} with $\alpha_\mrm{MLT} = 1.65$ to estimate the super-adiabatic temperature gradient $\Delta\nabla = \nabla - \nabla_\mrm{ad}$ at the point where the pressure $p = p_0\,\mrm{e}^{-1/2}$, where $p_0$ is the pressure at the bottom of the convection zone. We then estimate the entropy contrast an adiabatic convective element would reach after having overcome approximately one pressure scale height in such an environment,
\begin{equation}
(\delta s)_\mrm{min} \equiv (\delta s)_\mrm{typ} = \frac{\mathbb{R}}{\nabla_\mrm{ad}}\Delta\nabla.
\label{eq:ds_min}
\end{equation}

The total mass settling rate $\dot{M}$ is scaled as (see Sect.~\ref{sect:overview})
\begin{equation}
\dot{M} = \dot{M}_0\,\frac{\dot{M}_\mrm{phot}}{2.04\times 10^{21}\,\mrm{g}\,\mrm{s}^{-1}} \left(\frac{(\delta s)_\mrm{typ}}{7.96\times 10^1\,\mrm{erg}\,\mrm{g}^{-1}\,\mrm{K}^{-1}}\right)^{-1},
\label{eq:mdot}
\end{equation}
where $\dot{M}_\mrm{phot}$ is the mass downflow rate at the point (close to the photosphere) where the downflows reach the maximum entropy contrast $(\delta s)_\text{max}$, and the constant $\dot{M}_0$ is adjusted until the solar model reproduces the observed Li abundance in the Sun. The numbers in the denominators in Eq.~\ref{eq:mdot} correspond to the current solar values. The downflow mass flux in the photosphere, $\mathcal{F}_\mrm{phot}$, is computed in the same way as $(\delta s)_\mrm{max}$ (Eq.~\ref{eq:ds_max}),
\begin{align}
\log\,\mathcal{F}_\mrm{phot} = b_0 + b_1 x + b_2 y,
\label{eq:f_phot}
\end{align}
where $b_0 = -1.475$, $b_1 = -0.239$, and $b_2 = 0.511$ in cgs units. No extrapolation is needed when using Eqs.~\ref{eq:ds_max} or \ref{eq:f_phot} with the stellar models considered in this work. The mass downflow rate $\dot{M}_\mrm{phot}$ is then
\begin{equation}
\dot{M}_\mrm{phot} = 4\pi R_*^2\,\mathcal{F}_\mrm{phot},
\label{eq:mdot_phot}
\end{equation}
where $R_*$ is the radius of the star.

\subsubsection{Computational approach}

We neglect the thermodynamic response of the star to the convective settling process. We do, however, estimate the convective flux that would be caused by this process to further constrain our model and check the plausibility of our assumptions (Sect.~\ref{sect:heat_flux}). All downflows considered pass through the bottom of the convection zone (as defined by the Schwarzschild criterion) and their entropy contrast with respect to their surroundings is the highest at that point, hence the convective flux reaches a maximum there. Its value relative to the total flux of energy is estimated to be
\begin{equation}
\hat{\mathcal{F}}_\mrm{conv} = \frac{c_p T \overline{\frac{\Delta T}{T}} \dot{M}}{L},
\label{eq:f_conv}
\end{equation}
where $c_p$ is the heat capacity at constant pressure, $\overline{\Delta T/T}$ the mean temperature contrast in the distribution (weighted by the mass flow rate), and $L$ the luminosity of the star, all evaluated at the bottom of the convection zone. As explained in Sect.~\ref{sect:overview}, the mass flow rate for every downflow in our distribution corresponds to the point where the downflow settles, because we do not model mass entrainment explicitly. Most of the convective flux, however, is carried by the downflows that settle very close to the bottom of the convection zone, i.e. to the place where we compute $\hat{\mathcal{F}}_\mrm{conv}$. Hence, we regard Eq.~\ref{eq:f_conv} as a reasonable order-of-magnitude estimate.

To compute the Li and Be burning, we map the GARSTEC models (Sect.~\ref{sect:stellar_models}) on a grid equidistant in the mass fraction $q$.\footnote{Mass loss is negligibly small for the stars considered.} We only include the outermost 10\,--15\% of the stellar mass, which are relevant for the convective settling process. Interpolation of the models in time is done via the nearest-neighbour algorithm. We model the burning and transport of Li and Be using the set of equations
\begin{equation}
\frac{\diff A_i}{\diff t} = R_{\mrm{b},i} + R_{\mrm{m},i} + R_{\mrm{s},i} + R_{\mrm{a},i},
\label{eq:burning_equations}
\end{equation}
where $A_i = N_i/N_H$ is the abundance of Li or Be in the $i$-the grid cell, $N_i$ the number of Li or Be nuclei in the $i$-th grid cell, $N_H$ the number of hydrogen nuclei per grid cell,\footnote{There is no gradient in the hydrogen mass fraction, because we neglect the gravitational settling of He (see Sect.~\ref{sect:stellar_models}) and because convective settling does not reach the core of the star.} $i$ increases with radius, $t$ is the time, $R_{\mrm{b},i}$ the burning rate, $R_{\mrm{m},i}$ a mixing rate, $R_{\mrm{s},i}$ the settling rate, and $R_{\mrm{a},i}$ the advection rate due to the upflow. The burning rate
\begin{equation}
R_{\mrm{b},i} = -\frac{A_i}{\tau_{\mrm{b},i}}
\label{eq:burning_rate}
\end{equation}
is related to the nuclear-burning time scale $\tau_{\mrm{b},i}$, which we compute using the standard expressions for low-energy nuclear reaction rates in an ideal gas that can be found in  e.g.\ \citet{HansenKawaler94}. We consider the burning of $^7\mrm{Li}$ by the reaction $^7\mrm{Li}(\mrm{p},\,\alpha)\alpha$ and the burning of $^9\mrm{Be}$ by the reactions $^9\mrm{Be}(\mrm{p},\,\alpha)^6\mrm{Li}$ and $^9\mrm{Be}(\mrm{p},\mrm{d})^8\mrm{Be}$.\footnote{We omit the atomic mass numbers in the rest of the text.} The products of the last two reactions are quickly transformed to $^3\mrm{He}$ and $^4\mrm{He}$ nuclei, respectively, and are of no interest for this work. The low-energy astrophysical $S$-factors of the three reactions are taken from the NACRE-II database \citep{XuEtal13}. Electron screening is neglected. The burning time scale below the settling layer is set equal to that at the bottom of the layer for numerical reasons. The rate at which the downflows settle in the $i$-th grid cell is
\begin{equation}
\dot{m}_i = \dot{M}\left[F(s_{i+\nicefrac{1}{2}}) - F(s_{i-\nicefrac{1}{2}})\right],
\end{equation}
where the cumulative distribution function $F(s)$ is given by Eq.~\ref{eq:cumulative_distribution_function}, $s$ is the local entropy of the stratification, and the index $i+\nicefrac{1}{2}$ refers to the top and the index $i-\nicefrac{1}{2}$ to the bottom of the $i$-th grid cell. Every single cell of our equidistant grid contains a mass of $\Delta m$, so we can define a ``recycling'' time scale
\begin{equation}
\tau_{\mrm{r},i} = \frac{\Delta m}{\dot{m}_i},
\end{equation}
which is the time it takes the convective settling process to completely replace the content of the $i$-th grid cell by ``fresh'' material from the convection zone. Most of the downflows settle just below the convection zone and $\tau_{\mrm{r},i}$ can become very short there. To avoid severe time-step restrictions, we define a well-mixed zone, which is composed of the convection zone and the region where $\tau_{\mrm{r}} < 10^5$\,yr and $\tau_{\mrm{b}} > 10^8$\,yr. The fast recycling together with the slow burning prevent the formation of any significant gradients in the abundances of Li and Be in the well-mixed zone. We homogenise this zone using the artificial mixing term
\begin{equation}
R_{\mrm{m},i} = \frac{\overline{A}_\mrm{wmz} - A_i}{\Delta t},
\label{eq:mixing_rate}
\end{equation}
where $\overline{A}_\mrm{wmz}$ is the average abundance of Li or Be in the well-mixed zone and $\Delta t$ is the length of the current time step. We set $R_{\mrm{m},i} = 0$ outside the well-mixed zone. The downflows in our model bring the Li- and Be-rich material from the convection zone and deposit it in the settling layer. The settling rate of Li and Be  nuclei is
\begin{equation}
\dot{N}_{\mrm{s},i} = A_\mrm{cz}\frac{X\dot{m}_i}{m_\mrm{p}},
\end{equation}
where $A_\mrm{cz}$ is the abundance of Li or Be in the convection zone, $X$ the hydrogen mass fraction, and $m_p$ the proton mass. The rate of change of the abundance due to settling is then
\begin{equation}
\dot{R}_{\mrm{s},i} = \frac{\dot{N}_{\mrm{s},i}}{N_H} = A_\mrm{cz}\frac{\dot{m}_i}{\Delta m} = \frac{A_\mrm{cz}}{\tau_{\mrm{r},i}}.
\label{eq:settling_rate}
\end{equation}
Finally, there is an upflow due to mass conservation, which we model by the advection rate of Li or Be nuclei
\begin{equation}
\dot{N}_{\mrm{a},i} = A_{i-1}\frac{X\sigma_{i-\nicefrac{1}{2}}}{m_\mrm{p}} - A_{i}\frac{X\sigma_{i+\nicefrac{1}{2}}}{m_\mrm{p}},
\end{equation}
where $\sigma_{i-\nicefrac{1}{2}} = \sum_{k=0}^{i-1} \dot{m}_k$ is the mass inflow rate to the $i$-th grid cell at its bottom boundary and $\sigma_{i+\nicefrac{1}{2}} = \sum_{k=0}^{i} \dot{m}_k$ is the mass outflow rate from the $i$-th grid cell at its top boundary. The rate of change of the abundance due to advection is then
\begin{equation}
R_{\mrm{a},i} = A_{i-1}\sum_{k=0}^{i-1} \frac{1}{\tau_{\mrm{r,i}}} - A_{i}\sum_{k=0}^{i} \frac{1}{\tau_{\mrm{r,i}}}.
\label{eq:advection_rate}
\end{equation}
Equation~\ref{eq:burning_equations} is independent of the absolute abundance scale (see also Eqs.~\ref{eq:burning_rate}, \ref{eq:mixing_rate}, \ref{eq:settling_rate}, and \ref{eq:advection_rate}). Therefore, we start all our calculations with $A_i = 1$ and integrate Eq.~\ref{eq:burning_equations} using the standard, fourth-order Runge-Kutta method. The initial condition that we want to impose (see Sect.~\ref{sect:preliminaries}) is then taken care of simply by rescaling the results accordingly. We use the usual astrophysical notation $\log \epsilon = 12 + \log(A_\mrm{cz})$ in the rest of the text.

\section{Results}
\label{sect:results}

\subsection{Preliminaries}
\label{sect:preliminaries}

\begin{figure}
\centering
\includegraphics[width=9cm]{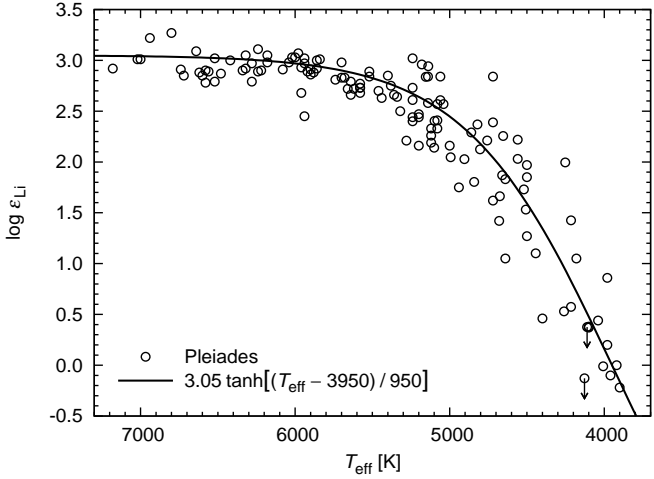}
\caption{Lithium distribution in the Pleiades \citep{SestitoRandich05}. The mean trend is approximated by a smooth function, which provides an initial condition for our model.}
\label{fig:initial_condition}
\end{figure}

Cool stars deplete significant amounts of their Li during their pre-main-sequence (PMS) evolution. Changes in the structure of a PMS star, however, are rather dramatic and show very large scatter even in a single cluster \citep{KenyonEtal90, KenyonHartman95, DunhamEtal08, BaraffeEtal09, EvansEtal09}. Applying our simple scaling relations to such a wide range of conditions would be questionable. Instead, we start all our computations at the age of the Pleiades, for which we adopt a value of 100\,Myr. We assume that the observed Li distribution in this cluster is a reasonable approximation to the PMS Li depletion. Ignoring the scatter at $T_{eff} = \text{const.}$, we model the mean trend by a smooth function (see Fig.~\ref{fig:initial_condition}) and use it as an initial condition for the Li abundance. The depletion of Be is much lower and difficult to measure at low $T_\text{eff}$, so we use the meteoritic value $\log\epsilon_\text{Be} = 1.30$ \citep{AsplundEtal09} as an initial condition for the Be abundance. We assume that both Li and Be are homogeneously distributed in the interior of the star at the start of the computation.

\begin{figure}
\centering
\includegraphics[width=9cm]{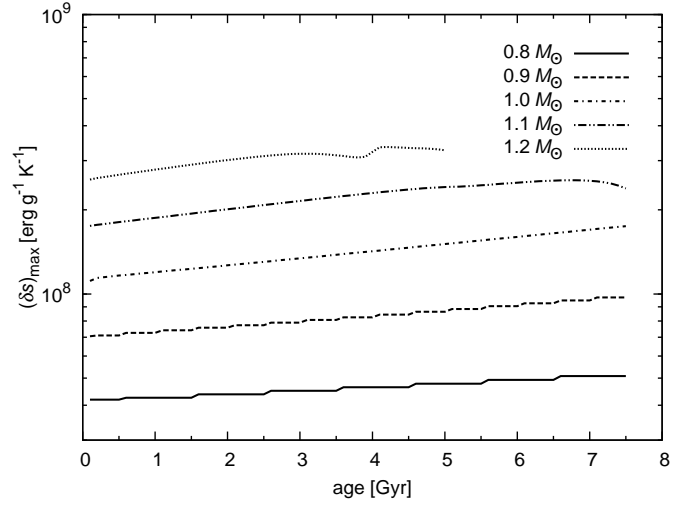}
\caption{Age dependence of $(\delta s)_\text{max}$ in the stellar models used in this work. The discontinuities are caused by the nearest-neighbour interpolation that we use.}
\label{fig:ds_max}
\end{figure}

\begin{figure}
\centering
\includegraphics[width=9cm]{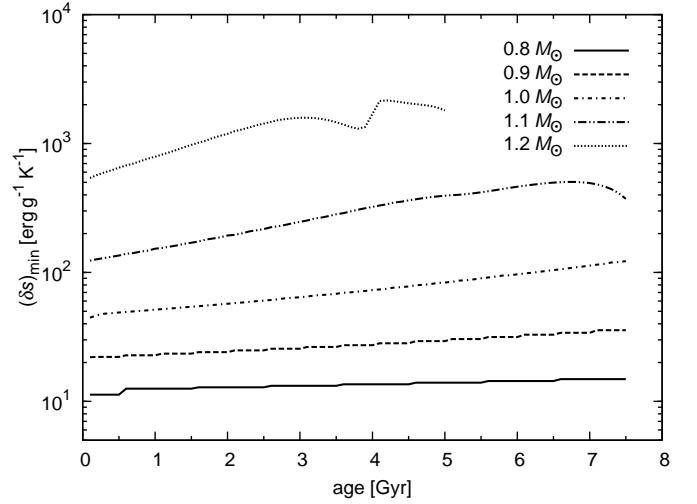}
\caption{As in Fig.~\ref{fig:ds_max}, but $(\delta s)_\text{min}$ is plotted.}
\label{fig:ds_min}
\end{figure}

\begin{figure}
\centering
\includegraphics[width=9cm]{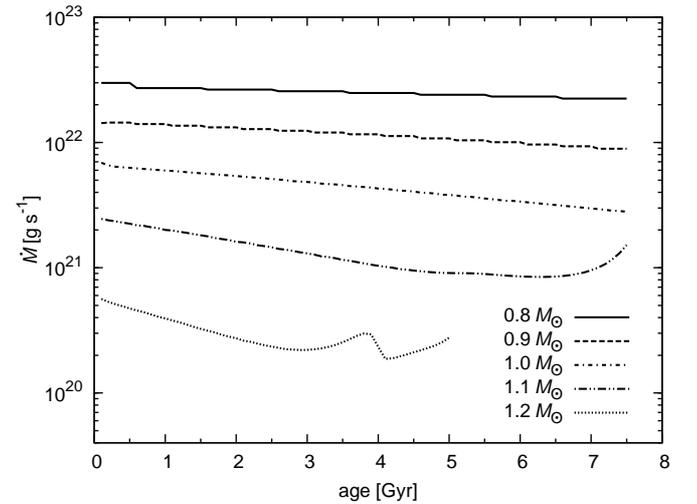}
\caption{As in Fig.~\ref{fig:ds_min}, but $\dot{M}$ is plotted. This figure assumes $\beta = 2.0$. The curves would be shifted downwards by $2.8$\,dex at $\beta = 1.5$ and upwards by $2.9$\,dex at $\beta = 2.5$.}
\label{fig:mdot}
\end{figure}

The dependences of $(\delta s)_\mrm{max}$, $(\delta s)_\mrm{min}$, and $\dot{M}$ on stellar mass and age, as defined by Eqs.~\ref{eq:ds_max}, \ref{eq:ds_min}, and \ref{eq:mdot}, are shown in Figs.~\ref{fig:ds_max}, \ref{fig:ds_min}, and \ref{fig:mdot}.
The values of $\dot{M}_\mrm{phot}$ in Eq.~\ref{eq:mdot} turn out to lie within $\sim 30\%$ of one another for all of the stars considered, although the spread in the values of $\mathcal{F}_\mrm{phot}$ in Eq.~\ref{eq:mdot_phot} is an order of magnitude larger. The scaling of $\dot{M}$ in Eq.~\ref{eq:mdot} is thus dominated by the $(\delta s)_\mrm{typ}$ factor.

\subsection{Li and Be depletion in the Sun}

\begin{figure}
\centering
\includegraphics[width=9cm]{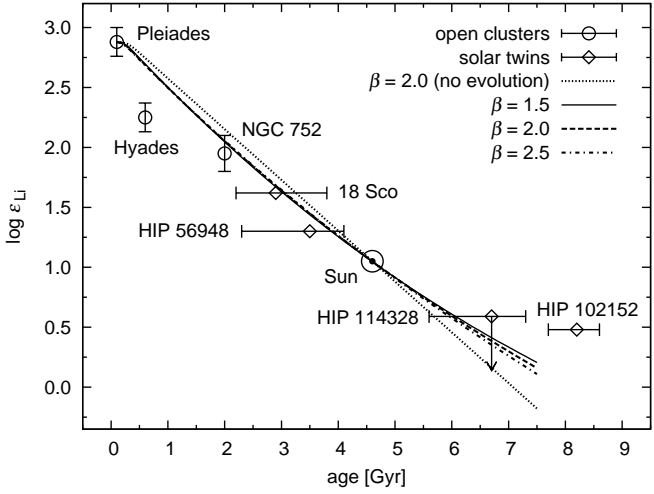}
\caption{Age dependence of the Li depletion in the Sun computed using our parametric model of convective settling compared with the Li abundances in four solar twins and in solar-type stars of three open clusters (the Sun is a calibration point and the Pleiades set the initial condition for Li in our model; see Sects.~\ref{sect:parameter_scaling} and \ref{sect:preliminaries}). The thermal stratification in the model plotted by the dotted line is not allowed to evolve in time. The abundances in the open clusters are from \citet{SestitoRandich05} and correspond to the solar effective temperature at the age of the cluster; the error bars show the typical scatter in the data. The measurements in 18~Sco and HIP~102152 are from \citet{MonroeEtal13}, and those in HIP~56948 and HIP~114328 from \citet{MelendezEtal12, MelendezEtal14}.}
\label{fig:solar_twins}
\end{figure}

The influence of stellar evolution on the Li depletion in the Sun is illustrated in Fig.~\ref{fig:solar_twins}. The depletion rate becomes quasi-stationary after $\sim 200$\,Myr and slowly decreases as the Sun ages. The depletion rate is hardly influenced by the assumed value of $\beta$. The observational data over-plotted in Fig.~\ref{fig:solar_twins} suggest a somewhat more pronounced slowdown in the depletion rate, although the error bars are quite large. We also show a non-evolving model, in which the stratification is given by the solar-structure model at an age of 4.6\,Gyr and is not allowed to evolve during the computation. In this case, the depletion rate becomes constant after the initial transition, as could be expected. 

The abundance of Be predicted by the evolving, $1.0\,M_\odot$ model at an age of 4.6\,Gyr ranges from 1.15 at $\beta = 1.5$ to 1.17 at $\beta = 2.5$, which deviates from the observed value of $1.38\pm 0.09$ \citep{AsplundEtal09} by $-2.5\sigma$. The meteoritic value is only $1.30\pm 0.03$ \citep{AsplundEtal09}.

\subsection{Mass dependence of Li and Be depletion}

\begin{figure}
\centering
\includegraphics[width=9cm]{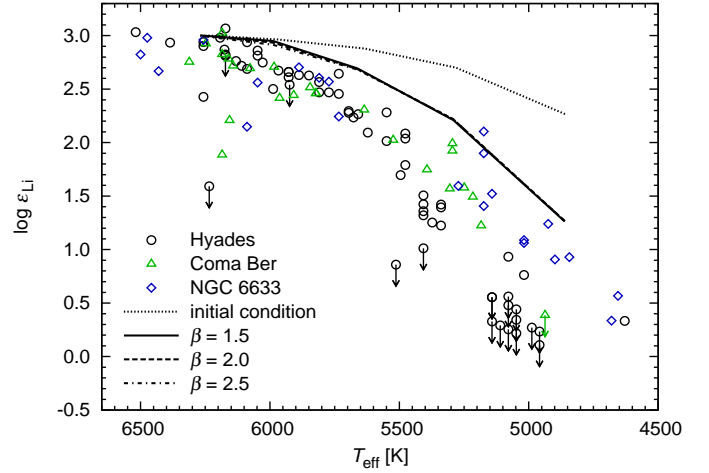}
\caption{Dependence on the effective temperature of the Li depletion computed using our parametric model of convective settling compared with the Li abundances in three 600-Myr-old open clusters as determined by \citet{SestitoRandich05}. Almost no dependence on $\beta$ can be seen at this age.}
\label{fig:clusters_0.6_gyr}
\end{figure}

\begin{figure}
\centering
\includegraphics[width=9cm]{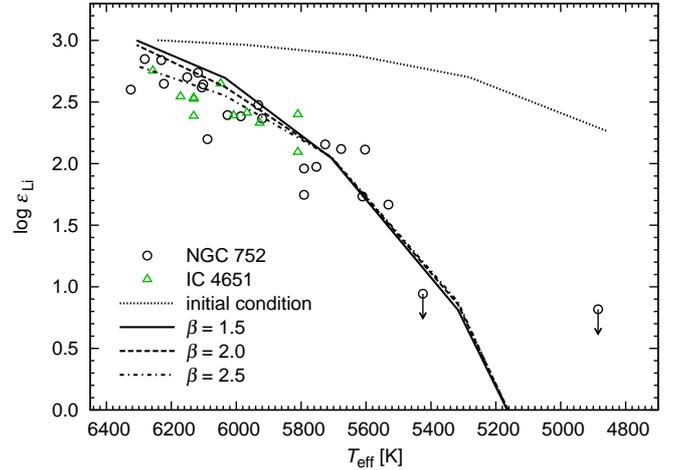}
\caption{As in Fig.~\ref{fig:clusters_0.6_gyr}, but the comparison is made with two 2-Gyr-old open clusters; the measurements are from \citet{SestitoRandich05}. A slight dependence on $\beta$ becomes visible at this age.}
\label{fig:clusters_2.0_gyr}
\end{figure}

Figures~\ref{fig:clusters_0.6_gyr} and \ref{fig:clusters_2.0_gyr} show the dependence of the Li depletion on the effective temperature of the star (hence on its mass) at a fixed age as compared with Li abundances observed in open clusters. The metallicity of the stars may influence the extent of Li depletion, because the higher the metallicity, the higher the opacity and the deeper the convection zone. The metallicities of the clusters used in this work are summarised in Table~\ref{tab:clusters}. Our model underestimates the Li depletion at an age of 600\,Myr (Fig.~\ref{fig:clusters_0.6_gyr}) independently of the value of $\beta$. This may stem from our assumption of a homogeneous Li distribution in the stellar interior at the start of the computation. The fit to the Hyades data (see Figs.~\ref{fig:solar_twins} and \ref{fig:clusters_0.6_gyr}) would also improve if the age of this cluster were not 600\,Myr as we assume, but 950\,Myr as \citet{BrandtHuang15} recently suggested. The data are scarce at an age of 2\,Gyr (Fig.~\ref{fig:clusters_2.0_gyr}), but the overall trend fits better. A weak dependence on $\beta$ can be seen at high effective temperatures.
\begin{table}
\centering
\begin{tabular}{c c c}
\hline\hline
Cluster\rule[-0.85ex]{0pt}{3.25ex} & Age [Gyr] & [Fe/H] \\
\hline
Pleiades\rule{0pt}{2.5ex} & 0.1 & $-0.03$ \\
Coma~Ber & 0.6 & $-0.05$ \\
Hyades & 0.6 & $+0.13$ \\
NGC~6633 & 0.6 & $-0.10$ \\
IC~4651 & 2 & $+0.10$ \\
NGC~752\rule[-0.85ex]{0pt}{2.5ex} & 2 & $+0.01$ \\
\hline
\end{tabular}
\caption{Ages and metallicities of the open clusters used in this work. Adopted from \citet{SestitoRandich05}; the age of the Pleiades is rounded to 0.1\,Gyr.}
\label{tab:clusters}
\end{table}

\begin{figure}
\centering
\includegraphics[width=9cm]{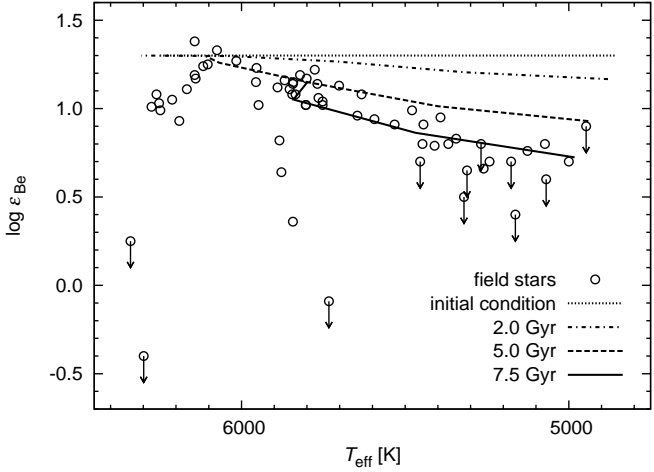}
\caption{Dependence on the effective temperature of the Be depletion computed using our parametric model of convective settling in a range of ages compared with the Be abundances measured in a sample of field stars from \citet{SantosEtal04}. The fastest evolving $1.2\,M_\odot$ star is computed up to 5~Gyr only.}
\label{fig:field_stars_be}
\end{figure}

The depletion of Be is much smaller and much more difficult to measure than the depletion of Li. Therefore, we resort to a comparison with field stars in Fig.~\ref{fig:field_stars_be}. We show models with \mbox{$\beta = 2.0$} only, because their dependence on $\beta$ is rather weak. The trend in the Be depletion is well reproduced assuming that the stars with $T_\text{eff} \lesssim 5500$\,K are older than $\sim 5$\,Gyr, which is a reasonable assumption for cool field stars.

\subsection{Heat flux due to convective settling}
\label{sect:heat_flux}

\begin{figure}
\centering
\includegraphics[width=9cm]{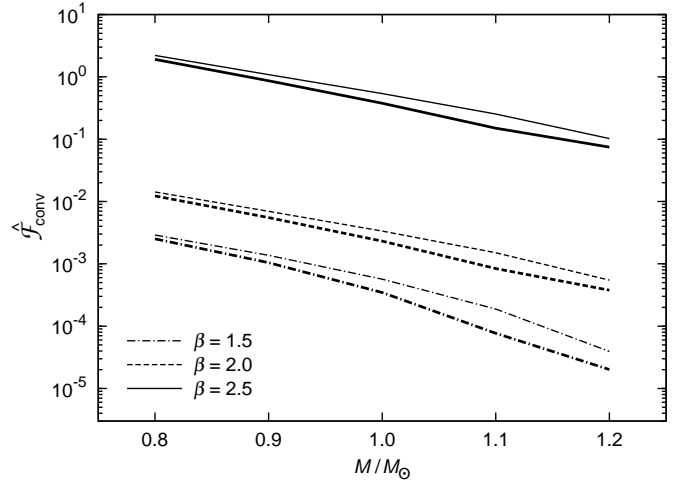}
\caption{Dependence on the stellar mass of the convective flux due to the convective settling process at the bottom of the convection zone. Three sets of models with $\beta \in \{1.5, 2.0, 2.5\}$ are shown at ages of 1~Gyr (thin lines) and 5~Gyr (thick lines).}
\label{fig:convective_fluxes}
\end{figure}

The convective flux at the bottom of the convection zone, $\hat{\mathcal{F}}_\text{conv}$ (Eq.~\ref{eq:f_conv}), unlike the depletion of Li and Be, is very sensitive to the assumed value of $\beta$, see Fig.~\ref{fig:convective_fluxes}. This comes about because the settling rate distribution spans several orders of magnitude in the downflow entropy contrast $\delta s$ and the slope $\beta$ of the distribution is constant over the whole range (see Sect.~\ref{sect:overview}). The calibration of the $1\,M_\odot$ model to the observed solar Li depletion sets the total amount of material that has to settle in the Li-burning layer. Thus, the calibration fixes the tail of the distribution where the entropy contrast is high, of the order of $(\delta s)_\text{max}$. If we increase the value of $\beta$ and recalibrate the model, the settling rate integrated over the Li-burning layer will not change much, but that just below the convection zone, where $\delta s \approx (\delta s)_\text{min} \ll (\delta s)_\text{max}$, will increase considerably. The main cause of the mass dependence of $\hat{\mathcal{F}}_\text{conv}$ seen in Fig.~\ref{fig:convective_fluxes} is the variation of $(\delta s)_\text{min}$ between different stars. This parameter changes $\hat{\mathcal{F}}_\text{conv}$ both directly, by changing the normalisation factor $N$ of the settling rate distribution (Eq.~\ref{eq:normalisation_factor}), and indirectly, via the scaling relation for $\dot{M}$, see Eq.~\ref{eq:mdot} and the related discussion in Sect.~\ref{sect:mathematical_formulation}. Since the values of $(\delta s)_\text{max}$ span a much narrower range than those of $(\delta s)_\text{min}$, the influence of this parameter on $\hat{\mathcal{F}}_\text{conv}$ is correspondingly weaker. The differences in the thermal stratification between the stars considered are significantly larger than those due to their main-sequence evolution. The dependence of $\hat{\mathcal{F}}_\text{conv}$ on the age of the star is therefore much weaker than on its mass (see Fig.~\ref{fig:convective_fluxes}). The models with $\hat{\mathcal{F}}_\text{conv}$ approaching or even exceeding unity are in conflict with our assumption of no thermal feedback of convective settling on the star (see the last paragraph of Sect.~\ref{sect:overview}), so their predictions should be taken with a grain of salt.

\section{Summary and discussion}

The process of convective settling is based on the idea that the envelope convection zones of low-mass, main sequence stars are dominated by large-scale downflows spanning the whole convection zone. If a tiny fraction of the low-entropy, photospheric downflows crosses the whole convection zone without having experienced much heating, their strongly negative buoyancy will make them sink and settle as deep as in the Li- and Be-burning layers. Mass conservation implies an upflow of the Li- and Be-depleted material back into the convection zone, reducing the photospheric abundances.

Building on the results of Paper~I, we have explored the dependence of the Li and Be depletion on stellar mass in the range from $0.8\,M_\odot$ to $1.2\,M_\odot$. We assume that the mass settling rate is distributed with respect to the entropy contrast of the downflow as a power law. It spans a wide range of entropy contrasts from $(\delta s)_\text{min}$, given by the MLT at the bottom of the convection zone, to $(\delta s)_\text{max}$, given by the maximum entropy contrast the downflows reach just below the photosphere. The slope $\beta$ of the distribution, assumed to be a constant, parametrises the physics of entrainment and heating processes acting on the downflows on their way to their settling points. In the absence of information on the dependence of the total mass settling rate $\dot{M}$ on the structure of the star, we scale this parameter in proportion to the mass downflow rate in the photosphere and in inverse proportion to $(\delta s)_\text{min}$, so that we qualitatively capture the effect of radiative diffusion on the heating of downflows. The mass downflow rate in the photosphere turns out to be essentially constant for the stars considered. We calibrate the scaling of $\dot{M}$ so as to reproduce the solar Li depletion. We allow for stellar evolution, but neglect the thermal feedback of convective settling on the star because we have shown in Paper~I that the feedback is negligible provided that $\beta \leq 2.5$. The computation is started at 100\,Myr, using the observed Li distribution in the Pleiades and the meteoritic Be abundance as initial conditions for Li and Be burning, respectively.

Changes in the solar structure cause a slowdown in the Li-depletion rate as the Sun ages. This slowdown seems to be somewhat milder in the model compared to the observed Li evolution in open clusters and solar twins. The main discrepancy occurs in the first $\sim 1$\,Gyr when real stars deplete Li faster than those in our model. This may be a consequence of our assumption that both Li and Be are homogeneously distributed in the star when the computation is started. Nevertheless, it is encouraging that the model can nicely reproduce the observed dependence of the Be depletion on the effective temperature in a sample of field stars. The current abundance of Be in the Sun predicted by the model is also compatible with the observed value. These conclusions are essentially independent of the assumed value of $\beta$, which is likely caused by the similarity of the internal structures of the stars considered.

We show that the convective flux at the bottom of the convection zone is very sensitive to $\beta$. The mass-flow-rate distribution in our model does not include the downflows with an entropy contrast lower than the MLT-based estimate $(\delta s)_\text{min}$. Such downflows carry a significant portion of the total flux even close to the bottom of the convection zone. Thus, one would expect the convective flux in our model to fall roughly into the interval $10^{-2} \lesssim \hat{\mathcal{F}}_\text{conv} \lesssim 10^0$ to be compatible with our conceptual picture of convective settling. This corresponds to $2.0 \lesssim \beta \lesssim 2.5$ in the solar model, in agreement with the conclusions of Paper~I.

The main caveat of our present analysis is the qualitative nature of the scaling of the total mass settling rate with the properties of the star. It determines the sensitivity of the predicted Li- and Be-depletion rates and of the convective flux on the stellar mass and age. We would need to model the details of the downflows' mass entrainment and heating to shed light on this issue, and to answer the question of whether the mass-settling-rate distribution is a power law in the first place.

\section*{Acknowledgements}

We thank Achim Weiss for the GARSTEC models used in this work, Zazralt Magic for the grid of model photospheres, and the anonymous referee for critical comments that improved the introduction section and the overall presentation of the text.

\end{document}